\documentstyle[12pt]{article}
\def\lqq{\lq \lq }

\begin{document}
\begin{titlepage}

\rightline{June 1995}
\rightline{IFUNAM FT 95-78, KL-TH 95/18 }
\begin{center}
{\LARGE  Exact Solvability of the Calogero and Sutherland Models}
\footnote{hep-th/9506105}
\vskip 0.3cm
{\large by
\vskip 0.3cm
Werner R\"uhl}$^{\it a}$\\
{\it Fachbereich Physik -- Theoretische Physik, University of Kaiserslautern,
Kaiserslautern D-67653, Germany}
\vskip 0.3cm
{\large Alexander Turbiner}$^{\it b}$\\
{\it Instituto de F\'isica, UNAM, Apartado Postal 20-364, 01000 Mexico D.F.,
Mexico}
\vskip 0.3cm
\end{center}

\begin{center}
{\large ABSTRACT}
\end{center}
\vskip 0.3 cm
\begin{quote}
Translationally invariant symmetric polynomials as coordinates for $N$-body
problems with identical particles are proposed. It is shown that in those
coordinates
the Calogero and Sutherland  $N$-body Hamiltonians, after appropriate gauge
transformations, can be presented as a {\it  quadratic} polynomial in the
generators
of the algebra $sl_N$ in finite-dimensional degenerate  representation.
The exact solvability of these models follows from the existence of the
infinite
flag of such representation spaces, preserved by the above Hamiltonians.
A connection with Jack polynomials is discussed.
\end{quote}

\noindent
$^{\it a}$E-mail: ruehl@physik.uni-kl.de \\[-0.1cm]

\noindent
$^{\it b}$On leave of absence from the Institute for Theoretical
and Experimental Physics,
Moscow 117259, Russia\\E-mail: turbiner@axcrnb.cern.ch,
turbiner@teorica0.ifisicacu.unam.mx
\end{titlepage}

The Calogero and Sutherland models are remarkable $N$-body classical and
quantum
problems with a pair-wise interaction that are completely integrable and at the
same time
are exactly solvable in their quantum versions \cite{calo, suth} (see also an
excellent
review \cite{olshper}). However, unlike the notion of integrability, which had
been
well established, for a long time the meaning of exact solvability remained
intuitive,
intrinsically folkloric. Just recently, an attempt to establish a certain
definition of
exact solvability was made \cite{ams}.

Let us take an infinite set of  finite-dimensional spaces $V_0, V_1,V_2,\ldots$
with an explicit basis. Assume that they are embedded one into another,
forming the infinite flag of spaces:
$V_0 \subset V_1 \subset V_2 \subset \ldots$.
 The simplest example of such a flag of spaces $V_n$ is given by a
linear space of polynomials in one variable of degree not higher than $n$ :
\begin{equation}
\label{e1}
V_n= \mbox{span} \{1, x, x^2, \ldots x^n \} \ .
\end{equation}
Then one can give a definition saying that {\it a  linear operator $h$ is named
{\bf exactly solvable} if it preserves the infinite flag }
of the spaces $V_n$ :
\begin{equation}
\label{e2}
h: V_n \mapsto V_n\ , \ n=0,1,2,\ldots
\end{equation}
Such a definition immediately leads to the conclusion that, in the basis
 where all spaces $V_n$ are naturally defined, the operator $h$ in matrix form
is given
by a triangular matrix. This does not contradict $h$ being self-adjoint.

Now one can state that our task is to find the flags of different spaces $V_n$,
where
the Hamilton operators act. Once we succeed in this task, we end up with the
{\it exactly solvable Hamilton operators}.

There are different ways to choose the spaces $V_n$. One of the most natural
choices
is to identify the space $V_n$  with a finite-dimensional representation space
of a certain
semi-simple Lie algebra. Then it is evident that an exactly solvable operator
$h$ generically should be a nonlinear combination of generators of the Cartan
and negative root
generators. For instance, the space (1) corresponds to the
$(n+1)$-dimensional
representation of the algebra $sl_2$ realized by first-order differential
operators acting
on the real line.

Of course, such a choice of $V_n$ is very particular. However,
it is quite amazing that all known one-dimensional exactly solvable quantum
mechanical
problems such as the harmonic oscillator, the Coulomb problem, etc., can be
interpreted in such
a manner as well as all exactly solvable discrete operators having Hahn,
Kravchuk,
Pollachek, Charlier and Meixner polynomials as their eigenfunctions
\cite{smtur}.
The purpose of the present paper is to show that the $N$-body Calogero and
Sutherland
models also belong to this type being matched  to finite-dimensional
representions
of the algebra $sl_N$. Thus the list of finite-dimensional exactly solvable
problems
 is exhausted and we arrived at the conclusion that all of them are of the same
type,
being presented by a quadratic combination of the generators of a certain Lie
algebra.

\noindent
{\it 1. The Calogero model}

The Calogero model \cite{calo} is a quantum mechanical system of $N$
particles on a line interacting via a pair-wise potential; it is defined by
the Hamiltonian
\begin{equation}
\label{e3}
H_{Cal} ={1 \over 2} \sum_{i=1}^N \left[ -d_i^2 + \omega^2 x_i^2 \right] +
\sum_{j < i}^N \frac g {(x_i-x_j)^2} \ \ \ \ \ ,
\end{equation}
where $d_i \equiv \frac{\partial}{\partial x_i}$, $\omega$ is the harmonic
oscillator frequency and hereafter normalized to $\omega=1$, and
$g=\nu(\nu -1)$ is the coupling constant.

As was found by Calogero \cite{calo}, this model is completely integrable and
possesses some very remarkable properties.
Firstly, in order to be normalizable, all eigenfunctions must  (up to a
certain function as a common factor) be either totally symmetric or totally
antisymmetric.
Secondly, it turned out that the energy spectrum is that of $N$
bosons or fermions interacting via harmonic forces only, but with a
total energy shift proportional to $\nu$ (for a review see \cite{olshper}).
Furthermore, the multiplicity of degeneracies of the states is the same as in
the
pure oscillator problem without repulsion. It is also worth noting that for
$N=3$
(and for this case only) the problem can be solved by a separation of
variables.

Thus the problem (\ref{e3}) is not only completely integrable, but also
exactly solvable:
the spectrum can be found explicitly, in a closed analytic form. All
eigenfunctions
have a form  ${\beta^{\nu^{\star}}} e^{-\frac{X^2}{2}} P(x)$, where $\beta$ is
the
Vandermonde determinant, $X^{2}=\sum_{i=1}^{N} {x_{i}}^{2}$, $ \nu^{\star}$
equals either $\nu$ or $(1-\nu)$, and
$P(x)$ is a completely symmetric polynomial under the permutation of any two
coordinates. Thirdly, after extracting the center-of-mass motion, the remaining
operator is translationally invariant under:  $x_i \rightarrow x_i +a$.

Now let us make a rotation of the Hamiltonian (\ref{e3}) \cite{tur} :
\[
h \equiv - 2 {\beta^{-\nu^{\star}}} e^{\frac{X^2}{2}} H_{Cal}
{\beta^{\nu^{\star}}}
e^{-\frac{X^2}{2}} =
\]
\[
=\  \sum_{i=1}^N  d_i^2 \ -\ 2 \sum_{i=1}^N  x_i d_i\ +\
\nu^{\star} \sum_{j\neq i}^N \frac 1 {x_i - x_j} [d_i - d_j]
\]
\begin{equation}
\label{e4}
- N - \nu^{\star} N(N-1) \ , \nonumber
\end{equation}
where $g=\nu^{\star}(\nu^{\star} - 1 )$, thereafter
the constant term in (4) will be omitted. Introduce the center-of-mass
\[
Y\ =\ \sum_{j=1}^N x_j\ ,
\]
and  translationally invariant  Jacobi coordinates \cite{per} (see also
\cite{tur}) :
\begin{equation}
\label{e5}
y_i\ =\ x_i - \frac{1}{N} \sum_{j=1}^N x_j\ ,\quad i=1,2,\ldots , N
\end{equation}
fulfilling the constraint $\sum_{i=1}^N y_i=0$.

It is known that the appropriate eigenfunctions of the operator (4) are
completely
symmetric under the permutations of $x$-coordinates \cite{calo} and,
consequently,
of $y$-coordinates. Now, instead of the explicit symmetrization of the
eigenfunctions of
(4), we encode this feature in a coordinate system consisting of functions in
the
$\{ x_{i} \}$ and $\{ y_{i} \}$, which are completely symmetric
under the permutations. The eigenfunctions, depending on these coordinates,
are then automatically symmetric.

There exist many different types of symmetric polynomials. For our purpose,
the most adequate set of symmetric polynomials is given by the elementary
symmetric
(ES) polynomials \cite{mac}:
\begin{eqnarray}
\label{e6}
\nonumber
\sigma_1(x) = \sum_{1 \leq i \leq N} x_i \ , \\
\nonumber
\sigma_2(x) = \sum_{1 \leq i < j \leq N} x_i x_j\ ,  \\
\nonumber
\sigma_3(x) = \sum_{1 \leq i < j < k \leq N} x_i x_j x_k\ ,  \\
\nonumber
\vdots \\
\sigma_N(x) = x_1 x_2 x_3\ldots  x_N \ ,
\end{eqnarray}
where $x \in \bf R$. We notice that $\sigma_1(x)$ coincides with the
center-of-mass
coordinate $Y$. Those polynomials can be defined also by a
generating function
\begin{equation} \label{a7}
   \prod^N_{i=1}  (1 + x_it) = \sum^N_{n=0} \sigma_n(x) t^n
\end{equation}
with $\sigma_0=1$.
The ES functions can be used as coordinates on the subset ${\bf E} \subset {\bf
R}^N$
\[
0< x_1 < x_2 < x_3 \dots < x_N < \infty
\]
of ${\bf R}^N$.  In turn, differential operators acting on ${\bf E}$ can be
rewritten in
terms of ES coordinates. If these differential operators are
themselves  symmetric, they can be extended from ${\bf E}$ to ${\bf R}^N$.

It is worth emphasizing that the coordinates (6) allow us to avoid the problem
of
over-completeness of the basis, which appears if the Newton polynomials
\begin{equation}
\label{a8}
s_k = \sum_1^N x_i^k
\end{equation}
are used as the coordinates \cite{per}.

Moreover, the Jacobian in the volume form
\begin{equation}
\label{a9}
d \sigma_{1} \wedge d \sigma_{2} \wedge \ldots \, \wedge d \sigma_{N}
= J_N dx_{1} \wedge dx_{2} \wedge \ldots \, \wedge dx_{N}
\end{equation}
is
\begin{equation}
\label{a10}
J_{N} = (-1)^{\left[ \frac{N}{2} \right]} \beta(x_{1}, x_{2}, \ldots \, ,
x_{N})
\end{equation}
with $\beta$ the Vandermonde determinant as before and  $[a]$ means the integer
part of $a$. In turn we can express $J_{N}^{2}$ in terms of the polynomials
$s_{k}$
(\ref{a8})
\begin{equation}
\label{a11}
J_{N}^{2} = \left| \begin{array}{ccccccc}
 N & s_1 & s_2 & s_3 & \ldots & & s_{N-1} \\
 s_1 & s_2 & s_3 & & & & s_N \\
 s_2 & s_3 & & & & & \\
 s_3 & & & & & & \\
 \vdots & & & & & & \\
 s_{N-1} & s_{N} & & & & & s_{2N-2}
 \end{array} \right|
 \end{equation}
These $s_k$ (including those with $k>N$) can be expressed by the $\sigma_{n}$
with the help of the generating function (\ref{a7})
\begin{equation} \label{a12}
\sum_{n=1}^{\infty} \frac{(-1)^{n+1}}{n} s_{n}(x) t^{n} = \log \left(
\sum_{n=0}^{N}
\sigma_{n}(x) t^{n} \right)\ .
\end{equation}

The coordinates (6) are not translationally invariant, but this property can
also be
included by considering the ES polynomials of the arguments (5),
\begin{equation}
\label{e7}
\tau_n (x) = \sigma_n(y(x)) \ , \ n=2,3,\ldots , N \ ,
\end{equation}
as the relative coordinates and with the remaining $\sigma_1(x)=Y$ as the
coordinate
of the center of mass. Certainly, once $n>N$, all $\tau_n (x) =0$. So $\tau_n$
implicitly
contains the information on the value of $N$.

Making quite sophisticated but straightforward calculations, one can get an
explicit
expression for the Laplacian in $\tau$-coordinates (7):
\begin{equation}
\label{e8}
\Delta \equiv \sum^N_{j=1} \frac{\partial^2}{\partial x_j^2}
 = N \frac{\partial^2}{\partial\sigma^2_1} +
	\sum^N_{j,k=2} A_{jk}
	\frac{\partial^2}{\partial\tau_j\partial\tau_k} +
	\sum^N_{i=2} B_{i}
	\frac{\partial}{\partial\tau_i} \ ,
\end{equation}
where
\[
	A_{jk} =  {(N-j+1) (k-1) \over N} \tau_{j-1} \tau_{k-1} +
	\sum_{\ell \geq {\rm max}(1,k-j)} (k-j-2\ell) \tau_{j+\ell-1}
	\tau_{k-\ell-1}
\]
\[
 	-B_i =   {(N-i+2) (N-i+1) \over N} \tau_{i-2}\ .
\]
Here we put $\tau_0=1, \tau_1=0$ and  $\tau_{p}=0$, if $p<0$ and $p>N$.

Finally, after the extraction of the center-of-mass motion and omitting the
constant
terms, the operator (4) reduces to $h_{rel}$ for the relative motion
\begin{equation}
\label{e9}
h_{rel} = \sum^N_{j,k=2} A_{jk}
	\frac{\partial^2}{\partial\tau_j\partial\tau_k}
-2 \sum^N_{i=2} i \tau_i { \partial \over \partial\tau_i} - \left(\frac{1}{N} +
\nu^{\star}\right)
 \sum^N_{i=2} (N-i+2)(N-i+1)  \tau_{i-2} { \partial \over \partial\tau_i }
\end{equation}

Now one can pose a question: would it be possible to rewrite (\ref{e9}) as
an element of the universal enveloping algebra of a certain Lie algebra in
a finite-dimensional representation?

Let us take the algebra $gl_N(\bf R)$. One of the simplest representations of
this
algebra in terms of first-order differential operators is the following :
\[
J_i^- = { \partial \over \partial t_i}  \ ,   \quad i=2,3,\ldots , N \ ,
\]
\[
J_{i,j}^0 = t_i J_j^-=t_i { \partial \over \partial t_j} \ , \quad
i,j=2,3,\ldots , N \ ,
\]
\[
J^0 = n - \sum_{p=2}^N t_p \frac{\partial}{\partial t_{p}} \ ,
\]
\begin{eqnarray}
\label{e10}
J_i^+ = t_i J^0\ , \quad i=2,3,\ldots , N \ .
\end{eqnarray}
which acts on functions of $t \in \bf R^{N-1}$. One of the generators, namely
$J^{0} + \sum_{p=2}^{N} J_{p,p}^{0}$ is proportional
to a constant and, if it is extracted, we end up with the algebra $sl_N(\bf
R)$.
The generators $J_{i,j}^0$ form the algebra of the vector fields $sl_{N-1}(\bf
R)$.
The parameter $n$ in (\ref{e10}) can be any real number. If $n$ is a
non-negative
integer, the representation (\ref{e10}) becomes the finite-dimensional
representation
acting on the space of polynomials
\begin{equation}
\label{e11}
V_n(t)\ =\ \mbox{span} \{ t_2^{n_2} t_3^{n_3} t_4^{n_4} \ldots t_{N}^{n_{N}} :
0 \leq \sum n_i \leq n\}\ .
\end{equation}
This representation corresponds to a Young tableau of one row and $n$ blocks
and
is irreducible.

It is easy to see that the operator $h_{rel}$ for the relative motion can be
rewritten in
terms of the generators (\ref{e10}).
The representation of $h_{rel}$ in terms of generators (\ref{e10}) is
\begin{eqnarray}
\label{e12}
h_{rel} &=& \sum_{j=2}^{N} \left\{ \frac{(N-j+1)(j-1)}{N} (J_{j-1,j}^{0})^{2} -
2 \sum_{\ell =1}^{j-1} \ell J_{j+\ell -1,j}^{0} J_{j-\ell -1,j}^{0} \right\}
\nonumber \\ &+&
2 \sum_{2 \leq k < j \leq N} \left\{ \frac{(N-j+1)(k-1)}{N} J_{j-1,j}^{0}
J_{k-1,k}^{0} -
\right. \nonumber \\
&-& \left. \sum_{\ell =1}^{k-1} (j-k+2\ell ) J_{j+\ell -1,j}^{0} J_{k-\ell
-1,k}^{0} \right\}
- 2 \sum_{k=2}^{N} k J_{k,k}^{0} \nonumber \\ &-&
\left(\frac{1}{N} + \nu^{\star}\right) \sum_{k=2}^{N}(N-k+2)(N-k+1)
J_{k-2,k}^{0}
\end{eqnarray}
where we identify $J_{0,k}^{0}$ with $J_{k}^{-}$ and put $J_{1,k}^{0}$ equal
to zero. Moreover the $t_{i}$ in (\ref{e10}) are replaced by the ${\tau}_{i}$.

Such a rewriting (\ref{e12}) can be performed when the parameter $n$ takes
any value. Hence, the operator $h_{rel}$ possesses infinitely many
finite-dimensional invariant subspaces $V_n(t),\ n=0,1,2,\ldots, $ and,
correspondingly,
preserves an infinite flag of the spaces $V_n(t)$. Therefore this operator is
exactly-solvable according to the definition given above. The case $N=3$
confirms
the general hypothesis stated in Ref.\cite{ams} about the non-existence of
exactly  and
quasi-exactly solvable problems in two-dimensional flat space without
separability
of variables. Also the operator (18) at $N=3$ is not contained in the lists
of the (quasi)-exactly-solvable operators in $R^2$ presented in the papers
\cite{qesr2}.

{\it 2. The Sutherland  model}

The Sutherland model \cite{suth} (for a review, see \cite{olshper})
is a quantum mechanical system of $N$ particles on a line interacting
via a pair-wise potential and defined by the Hamiltonian (cf. (\ref{e3}))
 ($d_{i} = \frac{\partial}{\partial x_{i}}$) :
\begin{equation}
\label{s1}
H_{Suth} = -\frac{1}{2} \sum_{i=1}^{N} d_{i}^{2} + \sum_{1 \leq i < j \leq N}
\frac{g}{{\sin}^{2} \mbox{$\frac{1}{2}$}
(x_{i} - x_{j})}
\end{equation}
which is defined on the Hilbert space of functions over the torus
$(S_{1})^{\times N}$. As in (3), $g=\nu(\nu -1)$ here is the coupling constant.

Similarly to the Calogero model, this model is completely integrable and
exactly-solvable -- the spectrum can be found explicitly, in a closed analytic
form. All eigenfunctions, in order to be normalizable,  must  (up to a
certain function as a common factor) be either totally symmetric or totally
antisymmetric and then have a form  ${\beta^{\nu^{\star}}} P(x)$, where
\[
\beta = \prod_{1 \leq i < j \leq N}  \left|\sin \frac{1}{2} (x_i-x_j)\right|
\]
is a natural modification of  the Vandermonde determinant,  $ \nu^{\star}$
equals
either $\nu$ or $1-\nu$,  while $P(x)$ is a completely symmetric polynomial in
coordinates $z=\exp {ix}$ under the permutation of any two coordinates.
These polynomials $P(z)$ are named Jack polynomials \cite{mac}.
It is evident that the Hamiltonian is translationally invariant under
$x_i \rightarrow x_i +a$.

Now we make the gauge transformation of the Hamiltonian (\ref{s1})
(see \cite{suth})
\begin{eqnarray}
\label{s2}
h &=& - 2 \beta^{- \nu^{\star}} H_{Suth} \beta^{ \nu^{\star}} \nonumber \\
&=& \sum_{i=1}^{N} d_{i}^{2} + i \nu^{\star} \sum_{1 \leq i < j \leq N} \cot
\mbox{$\frac{1}{2}$} (x_{i} - x_{j}) (d_{i} - d_{j}) + \mbox{const}
\end{eqnarray}
with the same relations between $g, \nu, \nu^{\star}$ as in the Calogero model.
The operator $h$ has Jack polynomials  as eigenfunctions.

In order to realize a possible hidden algebraic structure, one should introduce
a torus analogue of the symmetric polynomial coordinates (7). Let us consider
the torus
\[
 	T_N = (S_1)^{\times N}
\]
with standard coordinates
\begin{equation}
\label{s3}
  z_i (x) = e^{ix_i}, \qquad i =1, \ldots N \ ,
\end{equation}
\[
0 \leq x_i < 2\pi \ .
\]
Introduce the ES polynomials (see (6))
\[
\xi_n(x)  =  \sigma_n (z(x)) \ ,
\]
\begin{equation}
\label{se2}
\eta_n(x) = \sigma_n (w) \ ,
\end{equation}
where
\begin{equation}
\label{se3}
w_i  =  e^{iy_i} (x) , \qquad y_i = x_i  -  \frac1N \sigma_1 (x) \ .
\end{equation}
Obviously the coordinates (\ref{se2}) are translationally invariant
under $x_i \rightarrow x_i+a$ and
\begin{equation}
\label{se4}
\eta_n(x) = \frac{\xi_n(x)}{\bigg(\xi_N (x)\bigg)^{\frac{n}{N} }}  \ .
\end{equation}
Those variables possess a remarkable symmetry
\begin{equation}
\label{se5}
\eta^{\star}_n(x) = \eta_{N-n}(x)
\end{equation}
what implies that for even N the variable $\eta_{N/2}(x)$ is real.

Finally, as a complete system of coordinates we take
\begin{equation}
\label{s6}
\{ \xi_{N} \ , \eta_{n}, 1 \leq n \leq (N-1) \} \ ,
\end{equation}
where $\xi_{N}$ describes the motion of the center-of-mass.
In these coordinates the Laplacian has the form
\begin{eqnarray}
\label{s7}
- \Delta &=& N \left(\xi_{N} \frac{\partial}{\partial \xi_{N}}\right)^{2} +
\sum_{j,k=1}^{N-1}
 B_{jk} \frac{\partial^{2}}{\partial \eta_{j} \partial \eta_{k}} \nonumber \\
&+& \frac{1}{N} \sum_{l=1}^{N-1} l (N-l) \eta_{l} \frac{\partial}{\partial
\eta_{l}}
\end{eqnarray}
with
\begin{equation}
\label{s8}
B_{jk} = \frac{k(N-j)}{N} \eta_{j} \eta_{k} + \sum_{l \geq \max(1,k-j)}
(k-j-2l)
\eta_{j+l}\eta_{k-l}
\end{equation}
(cf. (\ref{e8})) and here $\eta_{0}=\eta_{N}=1$. The center-of-mass motion
can be separated and after omitting the constant, additive term
the Hamiltonian of the relative motion takes the form
\begin{eqnarray}
\label{s9}
- h_{rel} &=& \sum_{j,k=1}^{N-1} B_{jk} \frac{\partial^{2}}{\partial \eta_{j}
\partial \eta_{k}}\nonumber \\
&+& \left(\nu^{\star} + \frac{1}{N}\right) \sum_{l=1}^{N-1} l (N-l) \eta_{l}
 \frac{\partial}
{\partial \eta_{l}} \ .
\end{eqnarray}
With $t_{j}$ replaced by $\eta_{j-1}$ in (\ref{e10}) we obtain
\begin{eqnarray}
-h_{rel} &=& \sum_{j=1}^{N-1} \left\{ \frac{j(N-j)}{N} (J_{j,j}^{0})^{2} -
2 \sum_{l=1}^{j} l J_{j+l,j}^{0} J_{j-l,j}^{0} \right\} \nonumber \\
&+& 2 \sum_{1 \leq k < j \leq N-1} \left\{ \frac{k(N-j)}{N} J_{j,j}^{0}
J_{k,k}^{0} -
\sum_{l=1}^{k} (j-k+2l) J_{j+l,j}^{0} J_{k-l,k}^{0} \right\} \nonumber \\
&+& \nu^{\star} \sum_{l=1}^{N-1} l (N-l) J_{l,l}^{0}
\end{eqnarray}
Here we identified $J_{0,k}^{0}$ and $J_{N,k}^{0}$
with $J_{k}^{-}$. Like (18), the operator (30) at $N=3$
is not contained in the lists of the (quasi)-exactly-solvable operators in
$R^2$
presented in the papers \cite{qesr2}.

In conclusion, it is worth emphasizing that the Laplace operators (18), (27)
(from the physical point of view it can be treated such as we study a free
motion without an
interparticle interaction), after extraction of the center-of-mass motion,
are represented as a quadratic combination in the generators (\ref{e10})
of the algebra $sl_N(\bf R)$. The operator (29) realizes a new representation
of Jack polynomials in the coordinates (22). This representation is much
simpler than the standard representation (20) allowing easy an explicit
construction
of the first Jack polynomials.

\noindent
{\bf Acknowledgement }

\vspace{.2cm}
\noindent
This work was supported in part by the DFG grant 436 RUS 17/34/94 (Germany)
and the CONACyT research grant (Mexico).  A.T. wants to express his deep
gratitude for the kind hospitality extended to him at the University of
Kaiserslautern and the Theoretical Physics Department, Instituto de F\'isica,
UNAM, where part of the present work was done.

\end{document}